\documentclass[twocolumn,showpacs,superscriptaddress,amsmath,amssymb]{revtex4}
\usepackage{graphicx}
\usepackage{dcolumn}
\usepackage{bm}
\usepackage{color}

\def\bd{
\begin{document}} \def\ed{\end{document}}
\def\bmp{\begin{minipage}} \def\emp{\end{minipage}}
\def\bcc{\begin{center}} \def\ecc{\end{center}}     \def\npg{\newpage}
\def\beq{\begin{equation}} \def\eeq{\end{equation}} \def\hph{\hphantom}
\def\r#1{$^{[#1]}$}
\def\n{\noindent} \def\ni{\noindent} \def\pa{\parindent}
\def\hs{\hskip} \def\vs{\vskip} \def\hf{\hfill} \def\ej{\vfill\eject}
\def\cl{\centerline} \def\ob{\obeylines}  \def\ls{\leftskip}
\def\underbar#1{$\setbox0=\hbox{#1} \dp0=1.5pt \mathsurround=0pt
   \underline{\box0}$}   \def\ub{\underbar}    \def\ul{\underline}
\def\f{\left} \def\g{\right} \def\e{{\rm e}} \def\o{\over} \def\d{{\rm d}}
\def\vf{\varphi} \def\pl{\partial} \def\cov{{\rm cov}} \def\ch{{\rm ch}}
\def\la{\langle} \def\ra{\rangle} \def\EE{e$^+$e$^-$} \def\pt{p_{\rm t}}
\def\pti{p_{{\rm t},i}} \def\yti{y_{{\rm t},i}}
\def\ptj{p_{{\rm t},j}}\def\mt{m_{\rm t}} \def\yt{y_{\rm t}} \def\vt{v_{\rm t}}

\def\bitz{\begin{itemize}} \def\eitz{\end{itemize}}
\def\btbl{\begin{tabular}} \def\etbl{\end{tabular}}
\def\btbb{\begin{tabbing}} \def\etbb{\end{tabbing}}
\def\beqar{\begin{eqnarray}} \def\eeqar{\end{eqnarray}}
\def\\{\hfill\break} \def\dit{\item{-}} \def\i{\item}
\def\bbb{\begin{thebibliography}{9} \itemsep=-1mm}
\def\ebb{\end{thebibliography}} \def\bb{\bibitem}
\def\bpic{\begin{picture}(260,240)} \def\epic{\end{picture}}
\def\akgt{\cl{\bf ACKNOWLEDGMENTS}}
\def\fgn{\noindent{\bf\large\bf figure captions}}
\def\lan{\langle}
\def\ran{\rangle}
\def\mp1{\lan N_p\ran}
\def\map1{\lan N_{\bar p}\ran}
\def\Np1{N_p}
\def\Nap1{ N_{\bar p}}


\def\p{\pi}
\def\ifmath#1{\relax\ifmmode #1\else $#1$\fi}%
\def\rc{\ifmath{{\mathrm{c}}}}
\def\cut{\ifmath{{\mathrm{cut}}}}
\def\rF{\ifmath{{\mathrm{F}}}}
\def\rK{\ifmath{{\mathrm{K}}}}
\def\rp{\ifmath{{\mathrm{p}}}}
\def\rt{\ifmath{{\mathrm{t}}}}
\def\LAB{\ifmath{{\mathrm{LAB}}}}
\def\cut{\ifmath{{\mathrm{cut}}}}
\def\beq{\begin{equation}}
\def\eeq{\end{equation}}

\newcommand{\cinst}[2]{$^{\mathrm{#1}}$~#2\par}
\newcommand{\crefi}[1]{$^{\mathrm{#1}}$}
\newcommand{\crefii}[2]{$^{\mathrm{#1,#2}}$}
\newcommand{\crefiii}[3]{$^{\mathrm{#1,#2,#3}}$}
\newcommand{\HRule}{\rule{0.5\linewidth}{0.5mm}}

\bd
\title{Statistical and dynamical parts of the cumulants of conserved charges \\ in relativistic heavy ion collisions}

\author{Xue Pan}\email{panxuepx@gmail.com}
\affiliation{Key Laboratory of Quark and Lepton Physics (MOE) and
Institute of Particle Physics, Central China Normal University, Wuhan 430079, China}
\author{Fan Zhang}
\affiliation{Key Laboratory of Quark and Lepton Physics (MOE) and
Institute of Particle Physics, Central China Normal University, Wuhan 430079, China}
\author{Zhiming Li }
\affiliation{Key Laboratory of Quark and Lepton Physics (MOE) and
Institute of Particle Physics, Central China Normal University, Wuhan 430079, China}
\author{Lizhu Chen }
\affiliation{School of Physics and Optoelectronic Engineering, Nanjing University of Information Science and Technology, Nanjing 210044, China}
\author{Mingmei Xu }
\affiliation{Key Laboratory of Quark and Lepton Physics (MOE) and
Institute of Particle Physics, Central China Normal University, Wuhan 430079, China}
\author{Yuanfang Wu  }\email{wuyf@phy.ccnu.edu.cn}
\affiliation{Key Laboratory of Quark and Lepton Physics (MOE) and
Institute of Particle Physics, Central China Normal University, Wuhan 430079, China}

\begin{abstract}
The Poisson-liked statistical fluctuations, which are caused by the finite number of produced particles,
are firstly estimated for the cumulants of conserved charges, i.e., the cumulants of net-baryon,
net-electric charge, and net-strangeness. They turn out to be the same as those baselines
derived from Hadron Resonance Gas (HRG) model. The energy and centrality dependence of
net-proton cumulants at the Relativistic Heavy-Ion Collider (RHIC) are demonstrated to be mainly caused by statistical fluctuations. Subtracting the statistical fluctuations, the dynamical kurtosis of net- and total-proton from two versions of the
AMPT model and the UrQMD model at current RHIC beam energies are presented.
It is found that the observed sign change in the kurtosis of net-proton can not be reproduced by
these three transport models. There is no significant difference between net- and total-proton kurtosis in model
calculations, in contrary to the data at RHIC.
\end{abstract}


\maketitle
\section{Introduction}

The cumulants of conserved charges are suggested as good probes of Quantum Chromo Dynamics (QCD) phase boundary.
They are experimentally accessible and theoretically calculable.

At finite temperature and baryon chemical potential, effective chiral models~\cite{Asakawa-NPA504} and some lattice QCD
calculations~\cite{Fodor-CP, Karsch-CP} have predicted the existence of the QCD critical point (CP).
Since the higher order cumulants of conserved charges are more sensitive to the correlation length, they
are suggested as critical related measurements in heavy ion collisions~\cite{Stephanov-prl81,
Stephanov-prl91, Karsch-plb633, Stephanov-prl102, Stephanov-prl107}.

At vanishing chemical potential, the lattice QCD calculations
at physical quark masses have shown that the chiral crossover transition
appears as the remnants of the second order phase transition belonging to the $O(4)$ universality
class~\cite{Karsch-prd80, Gupta-prd85}. This makes it possible to explore the temperature of
QCD phase transition by the associated singularities of the higher order cumulants of conserved
charges~\cite{Karsch-plb695, Karsch-prd83, Bazavov-prd86, EMMI-Skokov}.

Moreover, recent calculations of lattice QCD indicate that the freeze-out conditions in heavy ion collisions
can be reliably determined by the ratios of the first three order cumulants of net-electric charge~\cite{Swagato-cpod13}.  So the
measurements of the cumulants of conserved charges are crucial in locating the QCD phase boundary.

Before understanding the physics of measured cumulants, the contributions of various non-critical effects should be figured out,
such as global conservation laws in a subsystem~\cite{Koch-Skokov}, initial size fluctuations~\cite{Xiaofeng12} and
experimental acceptance cuts~\cite{Koch-prc, Xiaofeng12}. In the paper,
we focus on the contributions of Poisson-liked statistical fluctuations, which are caused by the finite number of
produced particles~\cite{bialas, claude, kittel}.

For an ideal thermodynamic system, the number of particles is infinite. The statistical
fluctuations are small and negligible in comparison to the critical one. However, for the heavy ion collisions at RHIC, the number of produced particles is not infinite. For example, at the top energy of RHIC,
the mean of net-proton is less than 10~\cite{Star-prl}. Therefore, the statistical
fluctuations are not negligible.

The predictions of a multiphase transport (AMPT) model~\cite{AMPT} have shown that the behavior of the cumulants of net-proton is dominated by the statistical fluctuations at RHIC energies~\cite{Lizhu-JPG}. Here, the net-proton cumulants measured at experiment are further compared with corresponding statistical fluctuations directly. It shows clearly how the behavior of net-proton cumulants is dominated by the statistical fluctuations at nine centralities and three RHIC beam energies.

Subtracting the Poisson-liked statistical fluctuations, the dynamical cumulants of net-proton are recommended~\cite{Lizhu-JPG, Stephanov-prl107}. From the calculation of non-linear $\sigma$ model~\cite{Stephanov-prl107} and the arguments of universality near the critical point~\cite{Skokov-prc83}, the dynamical kurtosis of net-proton is negative when the critical point is approached from high temperature side.

The expected sign change has been observed in the corresponding experimental measurements at RHIC, i.e., the dynamical kurtosis of net-proton varies from negative to positive when the centrality varies from central to peripheral collisions, and the beam energy goes from high to low~\cite{Lizm-ismd12}. In contrast, the dynamical kurtosis of total-proton is positive at all beam energies and centralities. Whether the sign change in the dynamical kurtosis of net-proton indicates the appearance of critical point, or is simply caused by non-critical effects, or experimental cuts, is still not clear. A parallel investigation from known conventional models is helpful.

The paper is organized as follows. In section II, the statistical parts of the cumulants of three kinds of conserved charges are first derived, i.e., net-baryon, net-electric charge and net-strangeness. They turn out to be the same as the baselines derived from HRG model. Then the contributions of statistical fluctuations to the RHIC preliminary cumulants of net-proton are estimated in Section III. We find that the energy and centrality dependence of net-proton cumulants are dominated by the statistical fluctuations. In section IV, using the generators of the AMPT default, AMPT with string melting~\cite{AMPT}, and the Ultra Relativistic Quantum Molecular Dynamics (UrQMD) models~\cite{UrQMD}, the dynamical kurtosis of net- and total-proton at nine centralities and seven RHIC beam energies are presented, respectively. They are both positive, in contrary to the observed sign change in the dynamical kurtosis of net-proton of STAR data, but in consistent with the observed data of dynamical kurtosis of total-proton. Finally, the summary and conclusions are given in section V.

\section{ Statistical part of the cumulants}

As we know the statistical fluctuations of finite number of particles are well presented by the Poisson distribution~\cite{bialas}.
The possible charge carried by a baryon, an electric-charged particle, and a strangeness are respectively 1, 1 or 2, and 1, or 2, or 3.
We start from the simplest case. Suppose the baryon ($N^B_1$) and the anti-baryon ($N^B_{-1}$) numbers both follow
the Poisson distribution. The probability distribution of net-baryon ($N_B=N^B_1-N^B_{-1}$) is therefore the cross-correlation of two Poisson
distributions, i.e.,
\begin{equation}\label{Skellam}
\begin{split}
&f(N_B;\lan N^B_{1}\ran, \lan N^B_{-1}\ran)\\&=\sum_{x=-\infty}^{\infty}f(N_B+x,\lan N^B_1 \ran)f(x,\lan N^B_{-1}\ran)
\\&=e^{-(\lan N^B_1 \ran+\lan N^B_{-1}\ran)}\sum_{x=-\infty}^{\infty} \frac{\lan N^B_1\ran^{N_B+x}\lan N^B_{-1}\ran^{x}}{x!(N_B+x)!}
\\&=e^{-(\lan N^B_1\ran+\lan N^B_{-1}\ran)}(\lan N^B_{1}\ran/\lan N^B_{-1}\ran)^{N_B/2}I_{N_B}(2\sqrt{\lan N^B_{1}\ran\lan N^B_{-1}\ran}).
\end{split}
\end{equation}
Where $\lan N^B_1\ran$ and $\lan N^B_{-1}\ran$ are means of $N^B_1$ and $N^B_{-1}$, respectively. $I_{N_B}(z)$ is the modified Bessel function of the first kind. It is a standard Skellam distribution~\cite{skellam},  the same as that derived from HRG model~\cite{HRG}.

The cumulants of net-baryon ($\kappa_k^B$) can be obtained by the cumulant-generating function (CGF),
\begin{equation}\label{CGF-Skellam}
K_B(t; \lan N^B_1\ran, \lan N^B_{-1}\ran)=\sum_{k=0}^{\infty}\frac{t^k}{k!}\kappa_k^B.
\end{equation}
Where $K_B(t; \lan N^B_1\ran, \lan N^B_{-1}\ran)=\ln G(e^t; \lan N^B_1\ran, \lan N^B_{-1})\ran$, and $G(t;\lan N^B_1\ran,\lan N^B_{-1}\ran)$ is the probability-generating function (PGF) of Skellam distribution, i.e.,
\begin{equation}\label{PGF-Skellam}
\begin{split}
&G(t; \lan N^B_1\ran, \lan N^B_{-1}\ran)\\&=\sum_{N_B=0}^{\infty}f(N_B;\lan N^B_{1}\ran, \lan N^B_{-1}\ran)t^{N_B}\\&=G(t;\lan N^B_{1}\ran)G(1/t;\lan N^B_{-1}\ran)\\&=e^{-(\lan N^B_{1}\ran+\lan N^B_{-1}\ran)+\lan N^B_{1}\ran t+\lan N^B_{-1}\ran/t}.
\end{split}
\end{equation}

So the even and odd order cumulants of net-baryon are,
\beqar\label{even-odd-baryon}
\kappa_{2k}^B &=&\lan N^B_1\ran+\lan N^B_{-1}\ran,\nonumber  \\
\kappa_{2k+1}^B &=& \lan N^B_1\ran-\lan N^B_{-1}\ran.
\eeqar
They are uniquely determined by the means of baryon and anti-baryon numbers.

For electric charged particles, there are four kinds of particles, charge-one particle ($N_1^Q$) and anti-particle ($N_{-1}^Q$), and charge-two particle ($N_2^Q$) and antiparticle ($N_{-2}^Q$). Suppose the multiplicity of each kind of particles follows the Poisson distribution, the probability distribution of net-charge of charge-one particles ($N_{1Q}=N_1^Q-N_{-1}^Q$) is a Skellam distribution
again, the same as Eq.~{\eqref{Skellam}}. While, for charge-two particles, the probability distributions of charge ($2N_2^Q$) and anti-charges ($2N_{-2}^Q$) are not Poisson distribution, but
\beq\label{charge-two particle}
f(2N_2^Q; 2\lan N_{2}^Q\ran)=\lan N_2^Q\ran^{N_{2}^Q}e^{-\lan N_2^Q\ran}/N_2^Q!,
\eeq
and
\beq\label{charge-two anti-particle}
 f(2N_{-2}^Q; 2\lan N_{-2}^Q\ran)=\lan N_{-2}^Q\ran^{N_{-2}^Q}e^{-\lan N_{-2}^Q}\ran/N_{-2}^Q!,
\eeq
respectively. The probability distribution of the net-charge of charge-two particles  ($2N_{2Q}=2N_{2}^Q-2N_{-2}^Q$) is their cross-correlation,
\begin{equation}\label{net-charge2}
\begin{split}
&f(2N_{2Q};2\lan N_{2}^Q\ran, 2\lan N_{-2}^Q\ran)\\&=\sum_{x=-\infty}^{\infty}f(N_{2Q}+x,\lan N_2^Q\ran)f(x,\lan N_{-2}^Q\ran)
\\&=e^{-(\lan N_2^Q\ran+\lan N_{-2}^Q\ran)}\sum_{x=-\infty}^{\infty} \frac{\lan N_2^Q\ran^{{N_{2Q}}+x}\lan {N_{-2}^Q}\ran^{x}}{x!(N_{2Q}+x)!}.
\end{split}
\end{equation}
So, the probability distribution of the net-charge of all charged particles ($N_Q=N_{1Q}+2N_{2Q}$) is the convolution of
the probability distributions of the net-charges of charge-one and charge-two particles, i.e.,
\begin{equation}\label{netcharge}
\begin{split}
&f(N_Q;\lan N_{1}^Q\ran, \lan N_{-1}^Q\ran,2\lan N_{2}^Q\ran, 2\lan N_{-2}^Q\ran)\\&=\sum_{x=-\infty}^{\infty}f(x;\lan N_{1}^Q\ran,
\lan N_{-1}^Q\ran)f(N_Q-x;2\lan N_{2}^Q\ran, 2\lan N_{-2}^Q\ran)\\&= \sum_{x=-\infty}^{\infty}e^{-(\lan N_1^Q\ran+\lan N_{-1}^Q\ran)}
\sum_{n=-\infty}^{\infty} \frac{\lan {N_1^Q\ran}^{x+n}\lan {N_{-1}^Q}\ran^{n}}{n!(x+n)!}\\&~~~\times e^{-(\lan N_2^Q\ran+\lan N_{-2}^Q\ran)}
\sum_{m=-\infty}^{\infty} \frac{\lan {N_2^Q}\ran^{(N_Q-x)/2+m}\lan {N_{-2}^Q}\ran^{m}}{m!((N_Q-x)/2+m)!}.
\end{split}
\end{equation}

From its CGF and PGF, the even and odd order cumulants of net-charge can be derived easily,
\beqar\label{even-netcharge}
\kappa_{2k}^Q &=& \lan N_1^Q\ran+\lan N_{-1}^Q\ran+2^{2k}(\lan N_2^Q\ran + \lan N_{-2}^Q\ran), \nonumber\\
\kappa_{2k+1}^Q &=& \lan N_1^Q\ran - \lan N_{-1}^Q\ran + 2^{2k+1}(\lan N_2^Q \ran - \lan N_{-2}^Q\ran).
\eeqar
They are also consistent with those obtained from HRG model~\cite{HRG}. It should be noticed that the ratio $\kappa_4^Q/\kappa_2^Q$, i.e., the product of kurtosis and variance, is
\begin{equation}\label{ksigma-netcharge}
\frac{\kappa_4^Q}{\kappa_2^Q}=\kappa^Q_4\sigma^{2}_{Q}=\frac{\lan N_1^Q\ran + \lan N_{-1}^Q\ran+
16(\lan N_2^Q \ran+ \lan N_{-2}^Q\ran)}{\lan N_1^Q\ran + \lan N_{-1}^Q\ran +4(\lan N_2^Q \ran + \lan N_{-2}^Q \ran)}.
\end{equation}
It is not one, as the case of net-baryon, or net-proton~\cite{QM12-STAR-Daniel}, but determined by the means of the numbers of 4 kinds of charged particles.

For strangeness, there are six kinds of particles, i.e., strange one, two and three particles and antiparticles.
If the number of each kind of particles follows the Poisson distribution, similarly, the even and odd
order cumulants of net-strangeness can be derived,
\beqar\label{even-odd-strangeness}
\kappa_{2k}^S &=& \lan N_1^S\ran+\lan N_{-1}^S\ran+2^{2k}(\lan N_2^S\ran+\lan N_{-2}^S\ran) \nonumber\\
& & +3^{2k}(\lan N_3^S\ran+\lan N_{-3}^S\ran), \nonumber \\
\kappa_{2k+1}^S &=& \lan N_1^S\ran-\lan N_{-1}^S\ran + 2^{2k+1}(\lan N_2^S \ran - \lan N_{-2}^S\ran)\nonumber\\
& & +3^{2k+1}(\lan N_3^S\ran-\lan N_{-3}^S\ran) .
\eeqar
They are also the same as those obtained from HRG model~\cite{HRG}.

So starting from the assumption that all kinds of conserved charge particles and antiparticles are produced independently, or follow the Poisson distribution, the distributions and corresponding cumulants of the three kinds of net-charges are derived.
They turn out to be the same as those baselines obtained from HRG model, where the Boltzmann approximation is implemented and the quantum effects of electric-charged pion are neglected~\cite{HRG}. It shows that the baselines of the cumulants of conserved charges are in fact the fluctuations of independently produced particles, or pure Poisson-liked statistical fluctuations. They are completely determined by the means of particle and anti-particle numbers.

\section{Statistical part of the net-proton cumulants at RHIC }

The neutron is not detectable in experiments, however, the proton number fluctuations can reflect the singularity of baryon susceptibility very well~\cite{PRL91-Hatta}. Replacing the means of baryon and anti-baryon number in Eq.~\eqref{even-odd-baryon} by means of proton ($\mp1$) and anti-proton ($\map1$) numbers, the statistical cumulants of net-proton can be obtained. They are the variance,
\begin{equation}\label{variance}
\sigma_{p,stat}^2=\mp1+\map1,
\end{equation}
the normalized cumulants, i.e., skewness and kurtosis,
\begin{equation}\label{normalized cumulants}
S_{p,stat}=\frac{\mp1-\map1}{(\mp1+\map1)^{3/2}},\ \
\kappa_{p,stat}=\frac{1}{\mp1 +\map1},
\end{equation}
and the ratios of the third and fourth order cumulants to the second one, i.e., the products of skewness and standard deviation, and kurtosis and variance,
\begin{equation}\label{cumulant ratio}
S_{p,stat}\sigma_{p,stat}=\frac{\mp1-\map1}{\mp1+\map1},\ \
\kappa_{p,stat}\sigma_{p,stat}^2=1.
\end{equation}

From these expressions, it's clear that except $\kappa_{p,stat}\sigma_{p,stat}^2$ is a constant, all others are determined by the means of proton and anti-proton numbers. As we know, with increasing beam energy, and in more central collisions, the means of {\it produced} proton and anti-proton numbers increase. The difference between the means of proton and anti-proton becomes less and less. So the statistical parts of normalized cumulants and cumulant ratios decrease. In particular, $S_p\sigma_p$ decreases faster, and the skewness ($S_p$) is the fastest.

However, at lower beam energies, due to the baryon stopping effect~\cite{stopping}, a large number of protons from initial nuclei remain in formed system. It becomes more serious at even lower beam energies. For example, at 7.7 GeV, the lowest beam energy, the mean of anti-proton is two magnitudes smaller than that of proton. In the case, all statistical cumulants from Eq.~(4) are mainly determined by the mean of proton. This is why all order cumulants are close to each other at lower beam energies, cf., the Fig.~1 of ref.~\cite{Xiaofeng12}, and Fig.~2 of ref.~\cite{STAR-prl2}.

With increasing beam energy, the baryon stopping effect becomes weaker and weaker. At the higher beam energies, the contribution of anti-proton is not negligible. The statistical fluctuations of odd order cumulants is the mean of net-proton from Eq.~(4), and decrease rapidly with increasing beam energy. This is why the odd order cumulants decrease together, and obviously separate from even order ones with the increase of beam energy, cf., the Fig.~1 of ref.~\cite{Xiaofeng12}, and Fig.~2 of ref.~\cite{STAR-prl2}.

\begin{figure}[hbt]
\includegraphics[width=.48\textwidth]{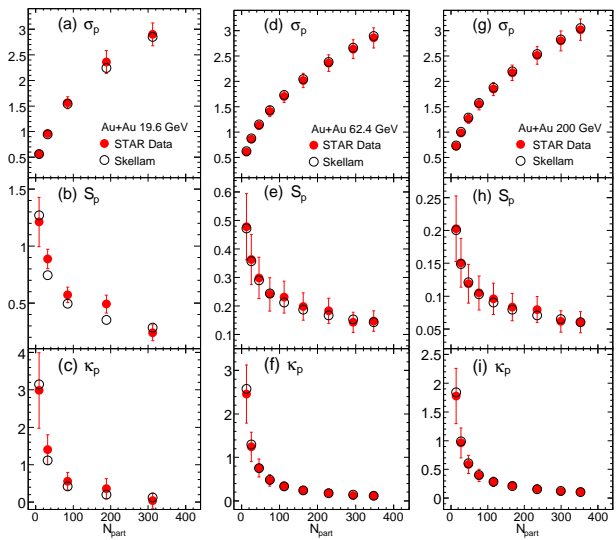}
\caption{\label{Fig. 1}(Color online) Centrality dependence of the statistical standard deviation ($\sigma_p$), skewness ($S_p$) and kurtosis ($\kappa_p$) of net-proton distribution (open black circles), and corresponding experimental data (solid red circles) for Au+Au collisions at $\sqrt{s_{NN}}$ = 19.6 GeV (left panel), 62.4 GeV (middle panel), and 200 GeV (right panel).}
\end{figure}

From the means of proton and anti-proton numbers at nine centralities and three RHIC beam energies, $\sqrt{s_{NN}}$ = 19.6, 62.4, and 200 GeV~\cite{Star-prl}, the statistical standard deviation, skewness and kurtosis of net-proton are calculated. They are presented by open black circles in Fig.~1, where the solid red circles are the data. The three panels from left to right correspond to the three beam energies, respectively.

The figure shows that all statistical cumulants (open black circles) are close to corresponding data (solid red circles).
The differences between them are one magnitude smaller. So the centrality and energy dependence of net-proton cumulants at RHIC
are dominated by its statistical parts. As expected, the statistical skewness $S_p$ is
firstly and greatly suppressed when the beam energy increases from 19.6 GeV to 200 GeV.

\section{Dynamical kurtosis of net- and total-proton at RHIC }

In order to see the difference between directly measured cumulants and statistical ones, the dynamical cumulants are recommended and defined as~\cite{Lizhu-JPG, claude},
\begin{equation}\label{dynamical cumulant}
\kappa_{p, dyn}=\kappa_p-\kappa_{p, stat}.
\end{equation}
It measures the correlations between charges. If the particles are produced independently, the dynamical cumulants
are zero.

The calculations from lattice QCD have shown that near critical temperature of chiral phase transition, the kurtosis at $\mu_B=0$ and $m_q=0$ is  positive. It could be negatively divergent near critical point at non-vanishing chemical potential and physical mass~\cite{Gupta}. The non-linear $\sigma$-model has demonstrated that if the critical point is approached from high temperature side, the dynamical kurtosis will change from negative to positive~\cite{Stephanov-prl107}, although the negative values are very small. The calculations of 3-dimensional Ising model show similar critical behavior~\cite{Panx-NPA}. While, the calculations of 3-dimensional $O(4)$ model show that the kurtosis oscillates between positive and negative. So the negative kurtosis is not specific to the critical end point (Ising universality). It may be associated with chiral phase transition ($O(4)$ universality)~\cite{EMMI-Skokov}. Anyway, the behavior of dynamical kurtosis at RHIC is highly interesting.

The dynamical kurtosis of net- and total-proton at nine centralities and seven RHIC beam energies ($\sqrt{s_{NN}}$ = 7.7, 11.5, 19.6, 27, 39, 62.4, and 200 GeV) are shown in Fig.~2(a) and 2(e),
respectively~\cite{Lizm-ismd12}. Fig.~2(a) shows clearly how the dynamical kurtosis of net-proton varies with two controlling parameters, i.e., energy and centrality. It is negative at non-central collisions and higher beam energies, i.e., $\sqrt{s_{NN}} > 19.6$GeV, and positive at $\sqrt{s_{NN}} < 19.6$ GeV. Where the black solid circles for the most peripheral collisions highlight this change. The positive and negative values in the dynamical kurtosis of net-proton indicate respectively that the peaks of net-proton distributions are sharper and flatter than those of corresponding Skellam distributions.

The figure also shows that the values of dynamical kurtosis are not zero, and one magnitude smaller than that directly measured. This indicates that proton and anti-proton are not independently produced at RHIC.

\begin{figure*}
\includegraphics[width=0.95\textwidth]{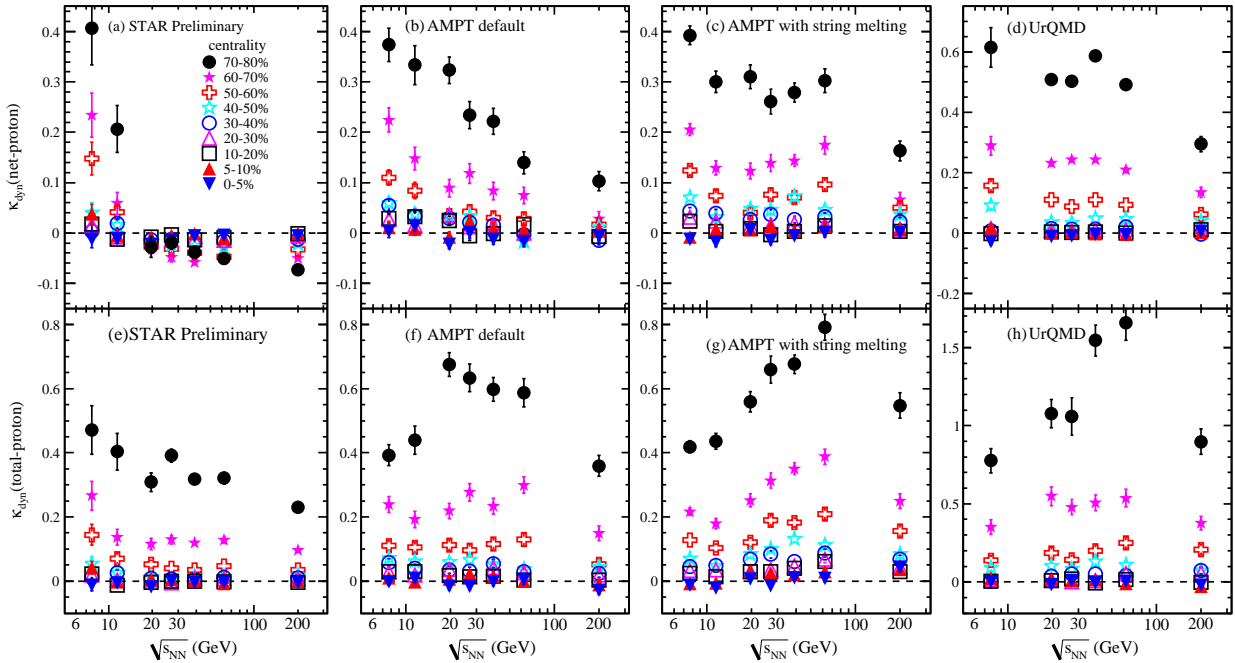}
\caption{\label{Fig. 2}(Color online) Energy dependence of the dynamical kurtosis of net-proton (upper panel) and total-proton (lower panel) at nine centralities for Au+Au collisions at RHIC. The results come from experimental data ((a) and (e))~\cite{Lizm-ismd12}, the AMPT default model ((b) and (f)), the AMPT with string melting model ((c) and (g)), and the UrQMD model ((d) and (h)), respectively. }
\end{figure*}

In order to see if the negative kurtosis can be caused by non-critical effects, or conventional particle production mechanisms,
we calculate the dynamical kurtosis in the AMPT default, the AMPT with string models~\cite{AMPT}, and the UrQMD model~\cite{UrQMD},  where no critical behavior is implemented in these models. As we know, the initial size fluctuations are well taken into account in these three transport models by Glauber model. However, the electric charge conservation of produced particles is not fully preserved in the AMPT models. As a compensation, the UrQMD model is better in taking the conservation of final state charges into account.

We simulate Au + Au collisions at seven corresponding beam energies by these three models. The calculations of dynamical cumulants of net-proton are performed in the same way as experimental analysis~\cite{Star-prl}. The centrality bins are selected by the multiplicity of charged particles except proton and anti-proton within pseudo-rapidity window $|\eta|<0.5$. The proton and anti-proton measurements are carried out at mid-rapidity window $|y|<0.5$ in the transverse momentum range $0.4<p_{T}<0.8 $ GeV/c. The dynamical kurtosis of net- and total-proton in each centrality bin are estimated by the Centrality Bin Width Correction (CBWC) method~\cite{Luoxf-JP}. They are presented in the upper and lower panels of Fig.~2, respectively. Where the first column is the data from RHIC/STAR. The second, third and fourth columns are the results from the AMPT default, the AMPT with string melting, and the UrQMD models, respectively.

From Fig.~2(b), (c) and (d), it's clear that the dynamical kurtosis of net-proton from the three model calculations are all positive at given centralities and energies, in contrary to the data in Fig.~2(a).
So the conventional particle production mechanisms implemented in these three transport models can not reproduce the observed
sign change in the dynamical kurtosis of net-proton. This inconsistency indicates that there should be additional correlations which has not been taken into account in these three transport models.

Whether is it critical related sign change? It's too early to make a conclusion. Since the absolute value of negative kurtosis is very small, less than 0.1, if the experimental cuts change, such as the phase space windows of the analysis, the definition of centrality and the centrality bin width corrections~\cite{Luoxf-JP}, the results may change accordingly. But, up to now, how to choose experimental cuts and how to reduce the non-critical effects are still in progress. So it is not ready yet for a conclusion.

On the other hand, the obtained results from current theoretical calculations and experimental measurements are encouraging. The behavior of dynamical kurtosis is very interesting, and worthwhile for the further investigations.

From Fig.~2(f), (g) and (h), it's also clear that the dynamical kurtosis of total-proton from the three model calculations are all positive, in consistent with data as shown in Fig.~2(e). They are all similar to those of dynamical kurtosis of net-proton in the model calculations as shown in Fig.~2(b), (c) and (d). So there is no significant difference between conserved and non-conserved charge in model calculations. However, the experimental data in Fig. 2(a) and (e) shows that the behavior of conserved charge is quite different from that of non-conserved charge. This indicates again that some correlations between conserved charges are missed in these three transport models.

The quantitative difference between two versions of the AMPT and the UrQMD models can be observed in the peripheral collisions. Where the results from the UrQMD model are all much larger than those from two versions of AMPT model. This may be caused by a strict conservation of final state charged particles in the UrQMD model. It leads to a stronger correlation between charged particles in peripheral collisions, where the number of produced particles are smaller than those of central collisions.

\section{Summary and conclusions}

In the paper, we argue that at RHIC beam energies, the Poisson-liked statistical fluctuations in higher order cumulants of
conserved charges are not negligible. Starting from independent particle production, i.e., assuming the Poisson
distribution for the number of conserved
charge particles, the statistical cumulants of net-baryon, net-electric charge, and net-strangeness are derived.
They are uniquely determined by the means of charged particle and anti-particle numbers, and
the same as those baselines obtained from HRG model. So the baselines
of higher order cumulants of conserved charges are essentially the statistical fluctuations.

From the means of proton and anti-proton numbers given by RHIC/STAR experiments, the
statistical standard deviation, skewness, and kurtosis are estimated. They are close to the data at nine centralities and three RHIC
beam energies. So the net-proton cumulants at RHIC are dominated by the statistical fluctuations.

Subtracting the statistical fluctuations, the dynamical kurtosis of net- and total-proton from two versions of the
AMPT and the UrQMD models at RHIC beam energies are presented. It is found that the dynamical kurtosis
of net-proton is small, but not zero. This indicates that proton and anti-proton are not produced independently
in these models, in consistent with data.

However, the observed sign change in the dynamical kurtosis of net-proton
at RHIC can not be reproduced by conventional particle production mechanisms implemented in these three models.
This inconsistency between the model calculations and experimental data indicates that there should be additional correlations between conserved charges in heavy ion collisions which has not been implemented in these three transport models.

In addition, model calculations show dynamical kurtosis of total-proton are all positive at observed centralities and energies,
in consistent with data. There is no significant difference between kurtosis of net- and total-proton,
or between conserved and non-conserved charges in the model calculations. However, from current experimental
data, the centrality and energy dependence of dynamical kurtosis of net-proton has a sign change, and the dynamical kurtosis of total-proton keeps positive. The behavior of dynamical kurtosis of net-proton is significantly different from that of total-proton.
It shows again that some correlation effects in conserved charges are missed in these models.

\section{Acknowledgement}

This work is supported in part by the NSFC of China with project No. 10835005, 11221504, 11005046, and 11005045, and the MOE of China for doctoral site with project No. 20120144110001.

\ed